\documentclass[12pt]{article}
\usepackage{a4wide,amsmath,amssymb,epsfig}
\begin{document}

\newcommand{\eqn}[1]{Eq.~\!(\ref{#1})}
\newcommand{\al}{\alpha}
\newcommand{\ga}{\gamma}
\newcommand{\la}{\lambda}
\newcommand{\si}{\sigma}
\newcommand{\de}{\delta}
\newcommand{\psib}{\bar{\psi}}
\newcommand{\vLa}{\varLambda}
\newcommand{\vTh}{\varTheta}
\newcommand{\vt}{\vartheta}
\newcommand{\ra}{\rightarrow}
\newcommand{\pr}{\prime}
\newcommand{\prop}{\mathcal{G}}
\newcommand{\twop}{\mathcal{C}}
\newcommand{\twoB}{\mathcal{B}}
\newcommand{\Dpath}{\mathcal{D}}
\newcommand{\Natu}{{\bf N}}
\newcommand{\Real}{{\bf R}}
\newcommand{\Comp}{{\bf C}}
\newcommand{\Kube}{{\bf K}}
\newcommand{\intk}{\int_\Kube}
\newcommand{\intkk}{\int_{\Kube^2}}
\newcommand{\half}{{\textstyle\frac{1}{2}}}
\newcommand{\sfrac}[2]{{\textstyle\frac{#1}{#2}}}
\newcommand{\Ord}{{\mathcal O}}
\newcommand{\Dp}{d\mu[\phi]}
\newcommand{\Exp}[1]{\mathsf{E}\!\left[#1\right]}
\newcommand{\diagram}[3]{\raisebox{-#3pt}{\epsfig{figure=#1.eps,
				                  width=#2pt}}}

\title{{\bf Quantum field theory for discrepancies II:\\ 
	    $\boldsymbol{1}\boldsymbol{/}\boldsymbol{N}$ corrections 
	    using fermions}}

\author{
Andr\'{e} van Hameren\thanks{andrevh@sci.kun.nl}~ and 
Ronald Kleiss\thanks{kleiss@sci.kun.nl}\\
University of Nijmegen, Nijmegen, the Netherlands\\
\and
Costas G. Papadopoulos\thanks{papadopo@alice.nuclear.demokritos.gr}\\
Institute of Nuclear Physics, NCSR 
$\Delta\eta\mu\acute{o}\kappa\varrho\iota\tau o\varsigma$, Athens, Greece}

\maketitle

\begin{abstract}
We calculate the $1/N$ corrections to the probability distributions of
quadratic discrepancies for sets of $N$ random points. This is achieved by the
introduction of fermionic variables. We give the diagrammatic expansion up to
and including the second order in $1/N$. For some discrepancies, we give the
explicit expansion to first order. 
\end{abstract}

\thispagestyle{empty}

\newpage
\pagestyle{plain}
\setcounter{page}{1}
\newpage

\section{Introduction}
{\sc Discrepancies} are measures of non-uniformity of point sets that play an
important r\^ole in the Quasi-Monte Carlo method of numerical integration
\cite{nieder1}. A certain class of these discrepancies, the so called {\em
quadratic discrepancies}, can be defined as an average-case complexity over a
class of functions, the {\em problem class} \cite{woz,kleiss,hak1}. In a number
of publications \cite{hk1,hk2,hk3} the problem of calculating the probability
distribution of discrepancies for sets of $N$ truly random points has been
tackled.  One of the results was the calculation of the asymptotic
distributions in the limit of infinite $N$. 

In \cite{hak1}, we introduced techniques from quantum field theory (QFT) to
calculate the moment generating function $G$ of the probability distribution,
suitable to calculate it as a series expansion in $1/N$. In this paper, we
extend the formalism by the introduction of fermions, and give the explicit
diagrammatic expansion of $\log G$ up to and including $\Ord(1/N^2)$. For the
Lego discrepancy, the $L_2^*$-discrepancy in one dimension and the Fourier
diaphony in one dimension, we give the explicit $1/N$ correction.

\section{The formalism}
We start with a short repetition of the formalism of \cite{hak1} and continue 
with the introduction of some new tools. 

\subsection{Quadratic discrepancies and path integrals}
We shall always take the integration region to
be the $s$-dimensional unit hypercube $\Kube=[0,1)^s$. The point set $X_N$
consists of $N$ points $x_k\in\Kube$, $k=1,2,\ldots,N$. 
Defined as an average-case complexity on the problem class of functions 
$\phi:\Kube\mapsto\Real$ with measure $\mu$, a discrepancy $D_N$ of the point
set $X_N$ is given by 
\begin{equation}
   D_N \;=\; N\int\eta^2_N[\phi]\,\Dp \quad,\quad
   \eta_N[\phi] \;=\; \frac{1}{N}\sum_{k=1}^N\phi(x_k) - \int_\Kube \phi(x)\,dx
   \;\;.
\label{CorEq001}   
\end{equation}
It is the quadratic integration error, averaged over the problem class.
The probability density $H$ of the discrepancy as a random 
variable of random point sets is calculated as the inverse Laplace transform of the moment generating function $G$:
\begin{equation}
   H(D_N=t) 
   \;=\; \frac{1}{2\pi i}\int\limits_{-i\infty}^{+i\infty}e^{-zt}G(z)\,dz 
   \quad,\quad
   G(z) \;=\; \Exp{e^{zD_N}} \;\;.
\label{CorEq026}   
\end{equation}
Here, $\mathsf{E}$ denotes the expectation value of a random variable.

The measure $\mu$ is assumed to be Gaussian and in \cite{hak1} it is shown that the generating function is therefore given by
\begin{equation}
   G(z) 
   \;=\; \int \left(\intk e^{g[\phi(x)-\intk \phi(y)\,dy]}\,dx\right)^N \Dp 
         \quad,\quad  g \;=\; \sqrt{\frac{2z}{N}} \;\;.
\label{CorEq002}
\end{equation}
In \cite{hak1} we suggested to put the $N^{\textrm{th}}$ power in the
exponential and interpret $G(z)$ as an Euclidean path integral 
\begin{equation}
   G(z) \;=\; \int \Dpath\phi\,\exp(-S[\phi])  
\end{equation}
with an action
\begin{equation}
   S[\phi] 
   \;=\; \frac{1}{2}\intkk \phi(x)\vLa(x,y)\phi(y)\,dxdy 
         - N\log\left(\intk e^{g[\phi(x)-\intk \phi(y)\,dy]}\,dx\right) \;\;,
\label{CorEq016}	 
\end{equation}
where $\vLa$ is the symmetric linear operator with boundary conditions 
which is the inverse of the two-point Green function under the measure $\mu$:
\begin{equation}
   \intk \vLa(x_1,y)\twop(y,x_2)\,dy
   \;=\; \de(x_1-x_2) \label{CorEq003} 
   \quad,\quad
   \twop(y,x_2) \;=\; \int\phi(y)\phi(x_2)\,\Dp \;\;.
\end{equation} 
The ``infinitesimal volume element'' $\Dpath\phi$ can be seen as defined by 
the rule that $\Dp=\Dpath\phi\,\exp(-S_0)$, where $S_0$ is the action 
with $z=0$. For notational convenience we put $\Dpath\phi$ to the left of the 
exponential.

One of the features of this formalism is that the action has a {\em gauge
freedom}; a {\em global translation} $\vTh_c:\phi(x)\mapsto\phi(x)+c$,
$c\in\Real$ leaves $\eta_N[\phi]$ and {\em a fortiori} $G(z)$ invariant, and
results in a change of the action at most linear in $\phi$:
\begin{equation}
   S[\vTh_c\phi]
   \;=\; S[\phi]+\al c\chi[\phi]+\half\al c^2 
   \quad \textrm{with} \quad
   \chi[\phi_1+\phi_2] \;=\; \chi[\phi_1]+\chi[\phi_2] \;\;.
\end{equation}
As a result of this, the path integral can be generally written as
\begin{equation}
   G(z) \;=\; \frac{1}{I[F]}\sqrt{\frac{2\pi}{\al}}\int\Dpath\phi\, 
               \exp(-F(\xi[\phi]) - S_\vTh[\phi])  \;\;,\quad 
   S_\vTh[\phi]=S[\phi]-\half\al\chi[\phi]^2 \;\;,
\end{equation}
where $F$ is restricted such that 
\begin{equation}
   I[F] \;\equiv\; \int_{-\infty}^{\infty} \exp(-F(c))\,dc
\end{equation}
exists, and $\xi$ is only restricted such that $\xi[\vTh_c\phi]=\xi[\phi]+c$.
It is, for example, possible to take $\chi[\phi]=\int_\Kube\phi(x)\,dx$ and
$F(\chi)=M\chi^2$ with $M\ra\infty$. In this gauge, $\chi[\phi]$ will vanish
from the action, which is reflected in a two-point function that integrates to
zero with respect to each of its variables. We shall refer to this gauge as the
{\em Landau} gauge. 
  
Clearly, the first equation of (\ref{CorEq003}) cannot be satisfied in the
Landau gauge, because then $\twop$ integrates to zero~\footnote{There is a
misprint in Eq.~(73) of \cite{hak1} with respect to this, where $\de(x-y)$
should be replaced by $\de(x-y)-1$.}. To see what happens, we assume that the
problem class is a vector space with a basis of at least square integrable
functions $\{u_n\}$, so that a member $\phi$ of the problem class can be
written as 
\begin{equation}
   \phi(x) 
   \;=\; \sum_n \phi_nu_n(x) \;\;,\quad \phi_n\in\Real \;\;,
\end{equation}
and that the Gaussian measure on this function space is defined by 
\begin{equation}
  \Dp 
  \;=\;
  \prod_{n}\frac{\exp(-\phi_n^2/2\si_n^2)}{\sqrt{2\pi\si_n^2}}\,d\phi_n
  \;\;,\quad \si_n\in\Real \;\;.
\label{CorEq004}  
\end{equation}
For the measure to be suitably defined, the strengths $\si_n$ have to
satisfy certain restrictions which can be translated into the 
requirement that $\Exp{D_N}$ exists. They are the inverse of the eigenvalues 
of $\twop$ and the basis consists of the eigenfunctions. Therefore, $\twop$ and 
$\vLa$ can be expressed in terms of the basis:
\begin{equation}
   \twop(x_1,x_2)
   \;=\; \sum_n\si_n^2u_n(x_1)u_n(x_2) \;\;,\quad
   \vLa(x_1,x_2)
   \;=\; \sum_n\frac{1}{\si_n^2}\,u_n(x_1)u_n(x_2) \;\;, 
\label{CorEq005}   
\end{equation}
where the last equation holds if the basis is orthonormal. The boundary
conditions satisfied by $\twop$ and $\vLa$ are those satisfied by the basis
functions. With different gauges come different bases and strengths. We call
a gauge in which the basis is orthonormal a {\em Feynman} gauge. If the
Landau gauge is used, in which $\intk\phi(x)\,dx=0$, then the basis functions
have to integrate to zero:
\begin{equation}
   \intk u^{(\textrm{L})}_n(x)\,dx \;=\; 0 \quad \forall\,n \;\;,
\end{equation}
where the label $\textrm{L}$ indicates the Landau gauge. This means that the 
basis cannot be ``complete'' in the sense that 
$\sum_nu_n(x)u_n(y)=\de(x-y)$, but that we have 
\begin{equation}
   \sum_n u^{(\textrm{L})}_n(x)u^{(\textrm{L})}_n(y) \;=\; \de(x-y) - 1 \;\;.
\end{equation}
The zero mode is isolated and integrated out of the path integral. We
want to stress that the gauge freedom is something that comes from the original
measure $\mu$, and that the Landau gauge exists for every quadratic
discrepancy. It is a result of the fact that an integration error is the same
for integrands that differ only by a constant.  

From now on we will denote the two-point function in the
Landau gauge by $\twoB$. It satisfies 
\begin{equation}
   \intk\vLa^{(\textrm{L})}(x_1,y)\twoB(y,x_2)\,dy \;=\; \de(x_1-x_2)-1\;\;, 
\end{equation}
and the discrepancy can then be written as 
\begin{equation}
    D_N \;=\; \frac{1}{N}\sum_{k,l=1}^N\twoB(x_k,x_l) \;\;. 
\end{equation}

\subsection{Fermions as tools to calculate the 
            $\boldsymbol{1}\boldsymbol{/}\boldsymbol{N}$ corrections}
In \cite{hak1} we suggested to make a straight forward expansion in $1/N$ of
$\exp(-S)$ to calculate $G$ perturbatively. This way, however, the calculation
of the perturbation series becomes very cumbersome, and the reason for this is
the following. We want to use the fact that an expansion in $1/N$ corresponds
to an expansion around $\phi=0$ of the part of the action that is non-quadratic
in $\phi$. The subsequent terms in the expansions are therefore proportional to
moments of a Gaussian measure, and can be calculated using diagrams (cf.
\cite{Rivers}). These diagrams, the {\em Feynman diagrams}, consist of lines
representing two-point functions and vertices representing convolutions of
two-point functions. Because the action is non-local, i.e. it cannot be written
as a single integral over a Lagrangian density because of the logarithm in
\eqn{CorEq016}, the total path integral, thus the total sum of all diagrams,
cannot be seen as the exponential of all {\em connected} diagrams, and it is
this that makes the calculations difficult.

In order to circumvent this obstacle, we introduce $2N$ Grassmann variables
$\psib_i$ and $\psi_i$, $i=1,2,\ldots,N$. They all anti-commute with each other
and commute with complex numbers:
\begin{align}
     \psib_i\psib_j + \psib_j\psib_i = 0 \quad&,\quad 
     \psib_i\psi_j + \psi_j\psib_i = 0 \quad,\quad 
     \psi_i\psi_j + \psi_j\psi_i = 0  \quad\quad i,j=1,2,\ldots,N \\
     c\psib_i - \psib_i c = 0 \quad&,\quad 
     c\psi_i - \psi_i c = 0 \quad\quad 
     i=1,2,\ldots,N \;\;,\quad c\in\Comp \;\;.
\end{align}
Now we use the well known Gaussian integration rules for Grassmann variables to 
write the $N^{\textrm{th}}$ power in \eqn{CorEq002} as an exponent and get 
\begin{align}
   & G(z) \;=\; 
     \int \Dpath\phi\,D\psib D\psi\, 
           \exp(-S[\phi,\psib,\psi])  \;\;, 
	   \\
   & S[\phi,\psib,\psi] \;=\;
     \frac{1}{2}\intkk \phi(x)\vLa(x,y)\phi(y)\,dxdy + 
     \intk e^{g[\phi(x)-\intk \phi(y)\,dy]}\,dx\sum_{i=1}^N\psib_i\psi_i \;\;,
     \label{CorEq006}
\end{align}
where we introduced the notational shorthand
\begin{equation}
    D\psib = d\psib_1d\psib_2\cdots d\psib_N \quad,\quad
     D\psi = d\psi_1d\psi_2\cdots d\psi_N  \;\;.
\end{equation}
Notice that this action contains the same gauge freedom, so that the action
becomes completely local if the Landau gauge is used. We have to introduce the
``fermion fields'' to achieve this, but for calculating the perturbation
expansion they are much easier to handle than the logarithmic potential. From
the action (\ref{CorEq006}) we obtain the Feynman rules. To calculate a term in 
the $1/N$-expansion of $G$, the contribution of all diagrams that can be drawn 
using the Feynman rules and carry the right power of $1/N$ has to be 
calculated.

Because in this paper we want to calculate the path integral itself, and no
correlation functions, we only have to consider {\em vacuum} diagrams, i.e.,
diagrams without external legs. Furthermore, we will always use the Landau
gauge, because then action is local, so that we only need to calculate {\em
connected} diagrams. The whole generating function is the exponential of the
sum of these diagrams. The contribution of the connected diagrams we denote by
$W$, so that 
\begin{equation}
   G(z) \;=\; e^{W(z)} \;\;.
\end{equation}
The diagrammatic expansion is an expansion in $1/N$, the subsequent terms of 
which we denote by 
\begin{equation}
   W(z) \;=\; W_0(z) + \frac{1}{N}W_1(z) + \frac{1}{N^2}W_2(z) + \cdots \;\;.
\end{equation}
In the Landau gauge the rules can be summarized by~\footnote{In \cite{hak1} we
called another two-point function the {\em propagator}, namely the inverse of
$\vLa(x,y)-2z\de(x-y)+2z$, denoted by $\prop_z(x,y)$. This one is obtained from
$\twoB(x,y)$ by the re-summation of certain diagrams, as we will show in the
following.}
\begin{align}
   &\textrm{boson propagator:}\hspace{6pt}\quad
    x\,\diagram{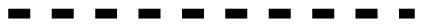}{50}{-1}\,y \;=\; \twoB(x,y) \;\;; \\
   &\textrm{fermion propagator:}\quad		  
    i\,\diagram{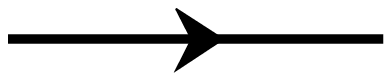}{50}{2}\,j \;=\; \de_{i,j}   \;\;;\\
   &\textrm{vertices:}\hspace{60pt}\quad
    \diagram{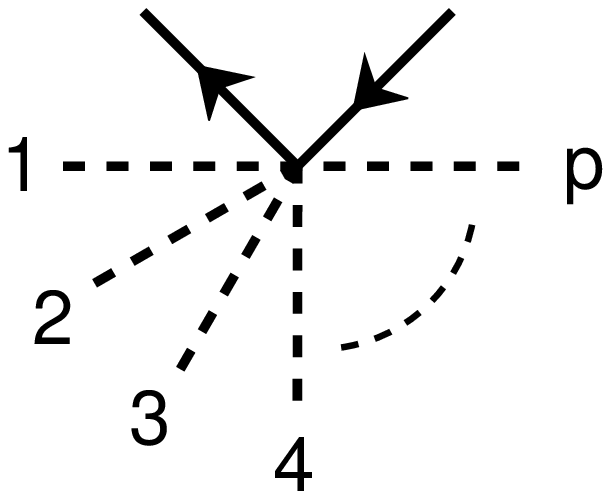}{65}{34}
    \;=\; -g^p\times\textrm{convolution} \;\;,\quad p\geq 2 
    \label{CorEq007}\;\;. 
\end{align}
In the vertices, boson propagators are convoluted as
$\intk\twoB(y,x_1)\twoB(y,x_2)\cdots\twoB(y,x_p)\,dy$, fermion
propagators as 
$\sum_{j=1}^{\raisebox{-1pt}{${\scriptstyle N}$}}\de_{i_1,j}\de_{j,i_2}$, 
and then these
convolutions are multiplied. As a result of this, the bosonic part of each
diagram decouples completely from the fermionic part. The contribution of the
fermionic part can easily be determined, for every fermion loop only gives a
factor $-N$. The main problem is now to calculate the remaining bosonic part.
Finally notice that, as a result of the Landau gauge, vertices with {\em only
one} bosonic leg do not exist.

\section{The diagrammatic expansion}
We want to stress again that we only need to calculate the connected
diagrams. The sum of the contributions of all these diagrams gives $W=\log G$.
Usually, a Feynman diagram is a mnemonic representing a certain contribution to
a term in a series expansion, i.e. a label. We will use the same drawing
for the contribution itself, apart of the symmetry factor of the diagram.
For example, the contribution of the diagram
$\diagram{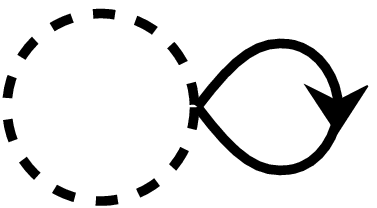}{30}{4.5}$ is equal to $\half Ng^2\intk\twoB(x,x)\,dx$ and its 
symmetry factor is equal to $\half$, so that we write 
\begin{equation}
   \frac{1}{2}\,\diagram{dcZ1}{40}{8} 
   \;=\; \frac{Ng^2}{2}\intk\twoB(x,x)\,dx \;\;.
\end{equation}

\subsection{The zeroth order}
The contribution to the zeroth order in $1/N$ can only come from diagrams in
which the power of $1/N$ coming from the vertices cancels the power of $N$,
coming from the fermion loops. This only happens in diagrams with vertices with
two bosonic legs only, and in which the fermion lines begin and end on the same
vertex. To write down their contribution, we introduce the two-point functions
$\twoB_p$, $p=1,2,\ldots$, defined by 
\begin{equation}
   \twoB_1(x_1,x_2) = \twoB(x_1,x_2) \;\;,\quad
   \twoB_{p+1}(x_1,x_2) = \intk\twoB_p(x_1,y,)\twoB(y,x_2)\,dy \;\;.
\end{equation}
The zeroth order term is given by 
\begin{align}
   \frac{1}{2}\,\diagram{dcZ1}{40}{8} \;+\; \frac{1}{4}\,\diagram{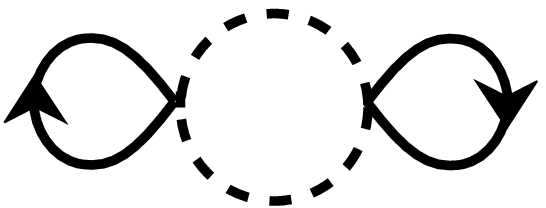}{60}{8} 
                       \;+\; \frac{1}{6}\,\diagram{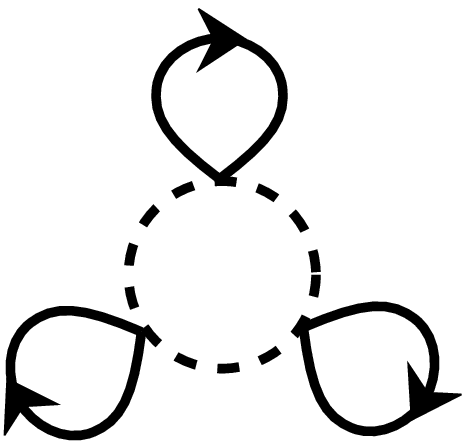}{50}{16} \;+\; \cdots 
   \;=\; \sum_{p=1}^{\infty}\frac{(Ng^2)^p}{2p}\intk\twoB_p(x,x)\,dx \;\;.
\end{align}
The factor $1/2p$ is the symmetry factor of this type of diagram with $p$
fermion ``leaves''. If we substitute $g=\sqrt{2z/N}$ in this expression, we
find exactly the result of Eq.~(21) in \cite{hkh1}. If we use the spectral
representation of $\twoB$ and assume that the basis functions are
orthonormal in the Landau gauge, we get 
\begin{equation}
   W_0(z) \;=\; -\frac{1}{2}\sum_n\log(1-2z\si_{\textrm{L},n}^2) \;\;,
\end{equation}
where $\textrm{L}$ indicates the Landau gauge. This expression is the 
same as Eq.~(68) in \cite{hk1}.

\subsection{The first order}
As we have seen before, bosonic two-point vertices with a closed single fermion
line contribute with a factor $2z$, and without any dependence on $N$.
Therefore, it is useful to introduce the following effective vertex
\begin{equation}
   \diagram{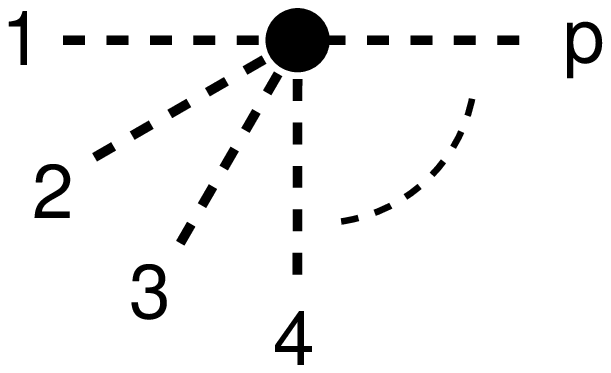}{65}{34} \;=\; \diagram{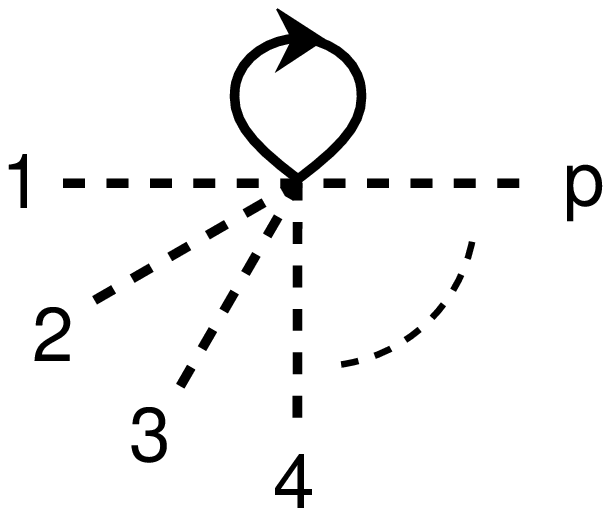}{65}{34}
                        \;=\; Ng^p\times\textrm{convolution}\;\;,
\end{equation}
and the following {\em dressed} boson propagator
\begin{align}
   x\,\diagram{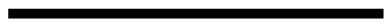}{50}{-1}\,y 
   \;&=\;       x\,\diagram{dcP2}{50}{-1}\,y \;+\; x\,\diagram{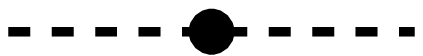}{50}{0}\,y 
          \;+\; x\,\diagram{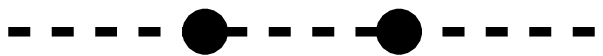}{75}{0}\,y \notag\\
     &\hspace{174pt}
	  \;+\; x\,\diagram{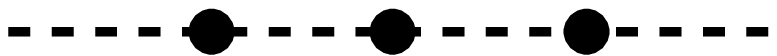}{100}{0}\,y \;+\; \cdots \notag\\
   \;&=\; \sum_{p=1}^\infty(2z)^{p-1}\twoB_p(x,y) \;\;.
\end{align}
If we assume that the basis is orthonormal in the Landau gauge, we can write 
\begin{equation}
   x\,\diagram{dcP6}{50}{-1}\,y 
   \;=\; \sum_n\frac{\si_{\textrm{L},n}^2}{1-2z\si_{\textrm{L},n}^2}\,
               u^{(\textrm{L})}_n(x)u^{(\textrm{L})}_n(y)\;\;,
\end{equation}
which is, apart of a factor $2z$, the same expression as in Eq.~(67) in
\cite{hk1}. The dressed propagator is equal to the propagator in the Landau
gauge as we defined it in Eq.~(24) in \cite{hak1}:
\begin{equation}
   x\,\diagram{dcP6}{50}{-1}\,y \;=\; \prop^{(\textrm{L})}_z(x,y) \;\;.
\end{equation}
This is easy to see, because it satisfies
\begin{equation}
   \intk\left[\vLa^{(\textrm{L})}(x_1,y)-2z\de(x_1-y)+2z\right]
        \sum_{p=1}^\infty(2z)^{p-1}\twoB_p(x,y)\;dy 
   \;=\; \de(x_1-x_2) - 1 \;\;,
\end{equation}
just like $\prop^{(\textrm{L})}_z$ by definition. Notice that 
$\prop^{(\textrm{L})}_z$ and $\twoB$ satisfy the relation 
\begin{equation}
   \lim_{z\ra0}\prop^{(\textrm{L})}_z(x,y)
   \;=\; \prop^{(\textrm{L})}_{z=0}(x,y) 
   \;=\; \twoB(x,y) \;\quad \forall\,x,y,\in\Kube  \;\;.
\end{equation}
Furthermore, notice that $\prop^{(\textrm{L})}_z$ and $W_0$ satisfy 
\begin{equation}
   \frac{\partial}{\partial z}\,W_0(z) 
   \;=\; \intk\prop^{(\textrm{L})}_z(x,x)\,dx \;\;,
\label{CorEq008}   
\end{equation}
and that this relation determines $W_0$ uniquely, because we know that 
$W_0(0)$ has to be equal to $0$ in order for the asymptotic probability 
distribution to be normalized to $1$.
From now on, we will omit the label $\textrm{L}$. 

The first order term in the expansion of $W(z)$ is 
\begin{equation}
   \frac{1}{N}W_1(z) 
   \;=\; \frac{1}{8}\,\diagram{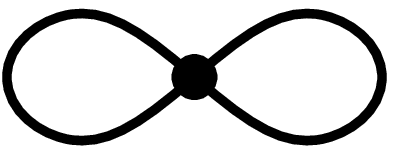}{40}{5}
   + \frac{1}{8}\,\diagram{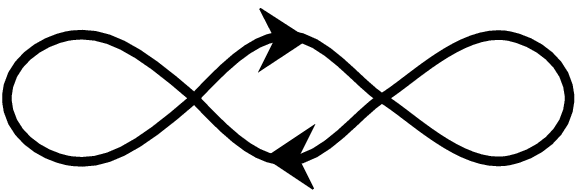}{60}{7}
   + \frac{1}{4}\,\diagram{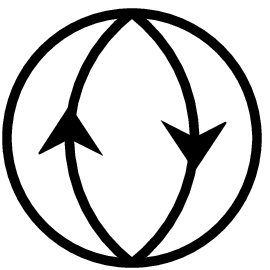}{30}{12}  
   + \frac{1}{8}\,\diagram{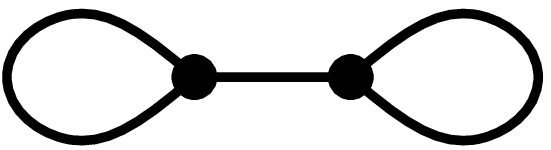}{58}{5}
   + \frac{1}{12}\,\diagram{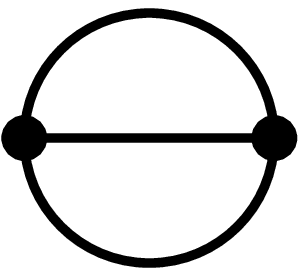}{30}{11} \;\;,
\label{CorEq009}
\end{equation}
or, more explicitly, 
\begin{align}
   W_1(z) 
   &\;=\;
   \frac{z^2}{2}\intk\prop_z(x,x)^2\,dx
             \;-\; \frac{z^2}{2}\left(\intk\prop_z(x,x)\,dx\right)^2
	     \;-\; z^2\intkk\prop_z(x,y)^2\,dxdy \notag\\ 
 &\hspace{22pt}+ z^3\intkk\prop_z(x,x)\prop_z(x,y)\prop_z(y,y)\,dxdy
             \;+\; \frac{2z^3}{3}\intkk\prop_z(x,y)^3\,dxdy  \;\;.
\end{align}

\subsection{The second order}
The second order term in the expansion of $W(z)$ is denoted by 
$\sfrac{1}{N^2}W_2(z)$ and is given by
\begin{align}
&    \frac{1}{48}\,\diagram{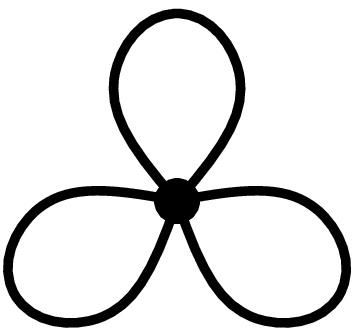}{35}{11}
\;+\;\frac{1}{48}\,\diagram{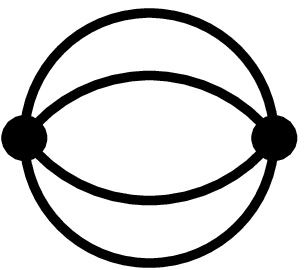}{30}{11}
\;+\;\frac{1}{16}\,\diagram{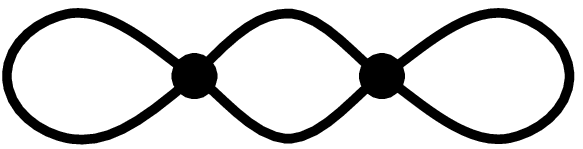}{60}{6}
\;+\;\frac{1}{12}\,\diagram{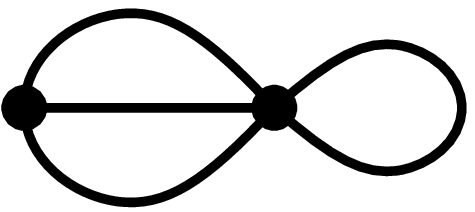}{50}{9}
\;+\;\frac{1}{24}\,\diagram{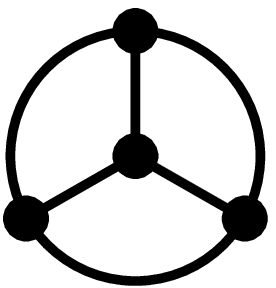}{30}{12}
\;+\;\frac{1}{16}\,\diagram{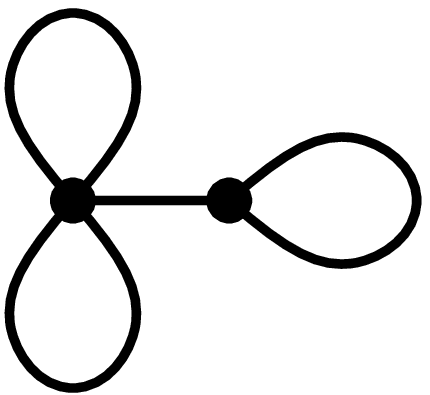}{45}{19}  \notag\\
&\;+\;\frac{1}{8}\,\diagram{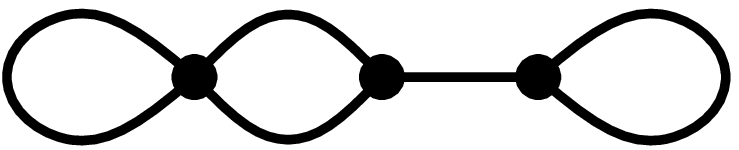}{75}{5}
\;+\;\frac{1}{8}\,\diagram{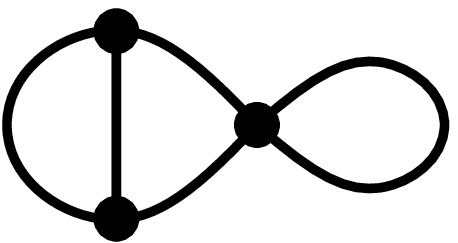}{50}{11}
\;+\;\frac{1}{8}\,\diagram{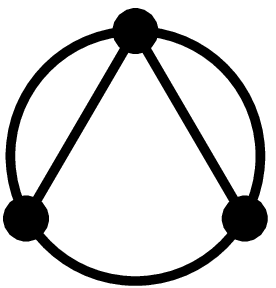}{30}{12}
\;+\;\frac{1}{16}\,\diagram{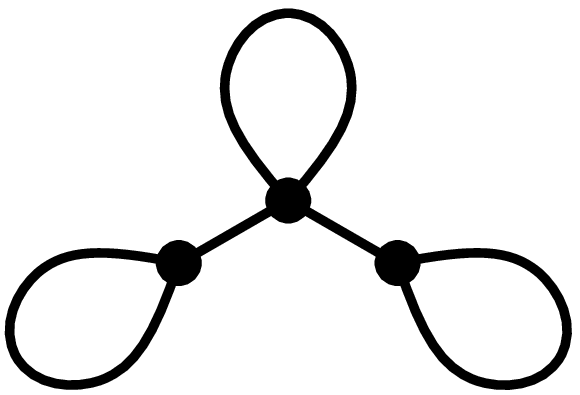}{60}{16}
\;+\;\frac{1}{12}\,\diagram{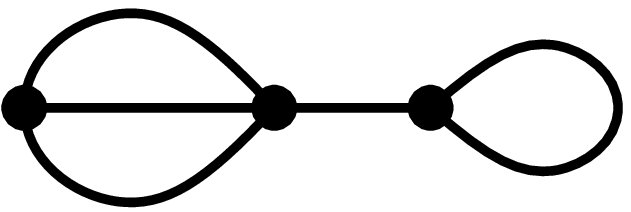}{65}{9}  \notag
\end{align}
\begin{align}
&\;+\;\frac{1}{48}\,\diagram{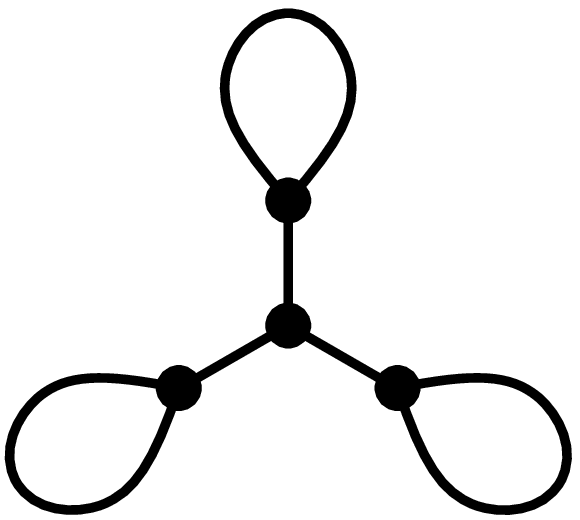}{60}{20}
\;+\;\frac{1}{8}\,\diagram{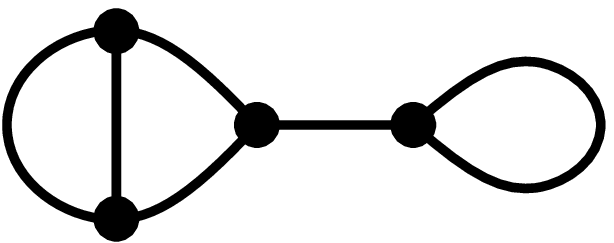}{65}{11}
\;+\;\frac{1}{16}\,\diagram{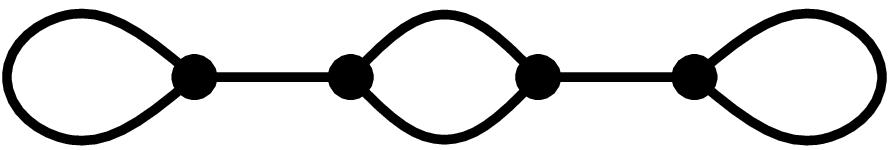}{90}{5}
\;+\;\frac{1}{16}\,\diagram{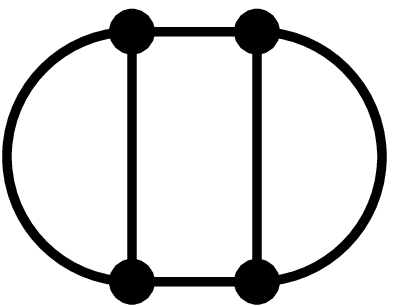}{40}{13} \notag\\
&\;+\;\frac{1}{4}\,\diagram{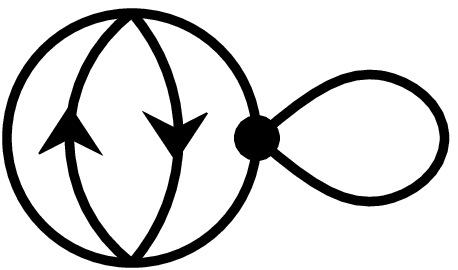}{50}{13}
\;+\;\frac{1}{8}\,\diagram{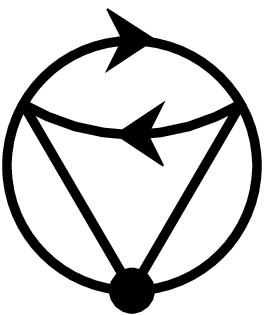}{30}{16}  
\;+\;\frac{1}{4}\,\diagram{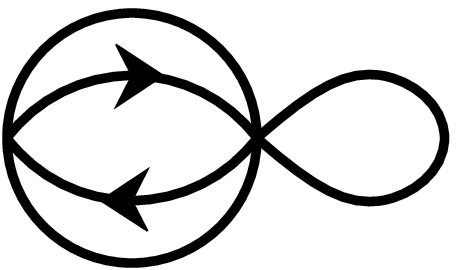}{50}{13}
\;+\;\frac{1}{4}\,\diagram{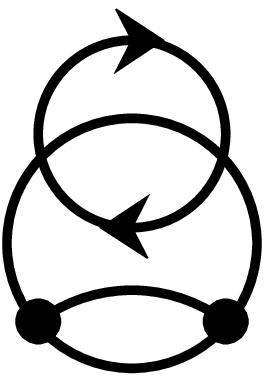}{30}{16}  
\;+\;\frac{1}{4}\,\diagram{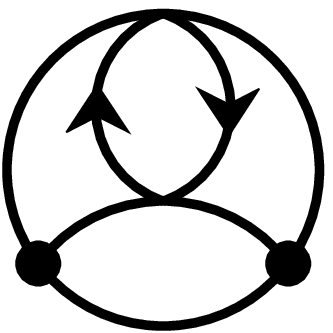}{30}{12}\notag\\
&\;+\;\frac{1}{2}\,\diagram{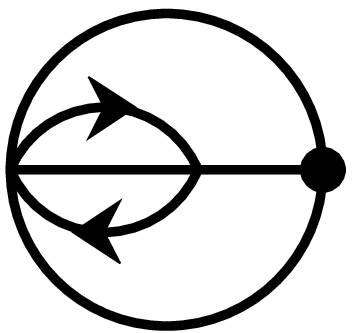}{30}{12}
\;+\;\frac{1}{16}\,\diagram{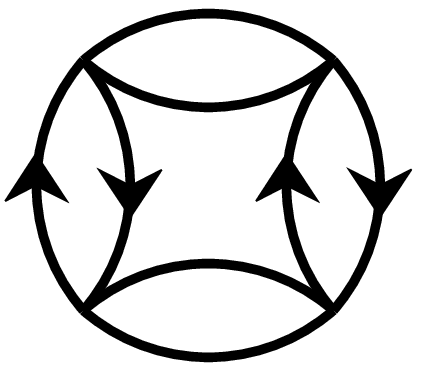}{40}{15}
\;+\;\frac{1}{8}\,\diagram{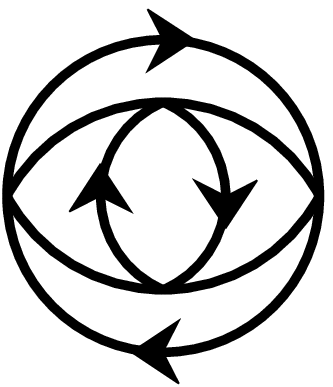}{35}{18}
\;+\;\frac{1}{4}\,\diagram{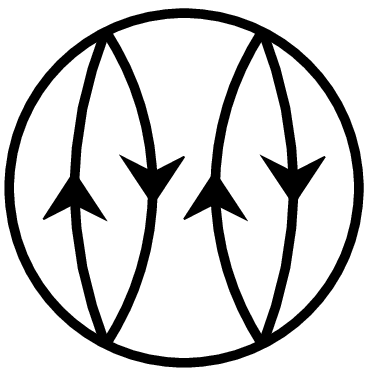}{35}{15}
\;+\;\frac{1}{12}\,\diagram{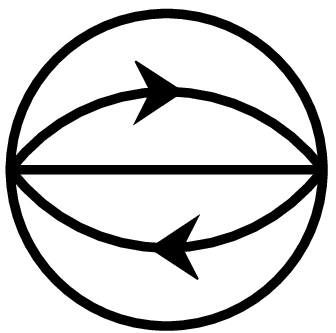}{30}{12} \notag\\
&\;+\;\frac{1}{8}\,\diagram{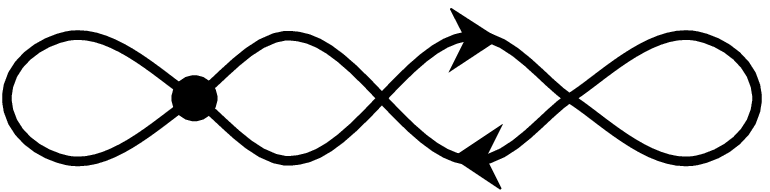}{80}{7}
\;+\;\frac{1}{16}\,\diagram{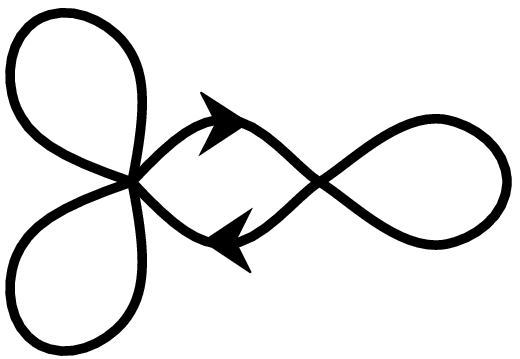}{55}{17}
\;+\;\frac{1}{8}\,\diagram{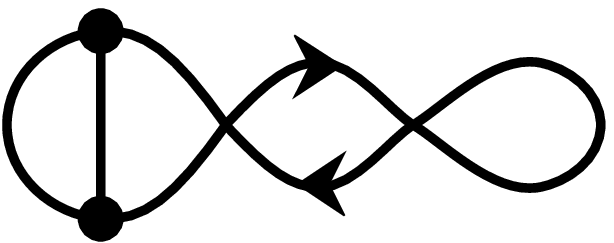}{65}{11}
\;+\;\frac{1}{12}\,\diagram{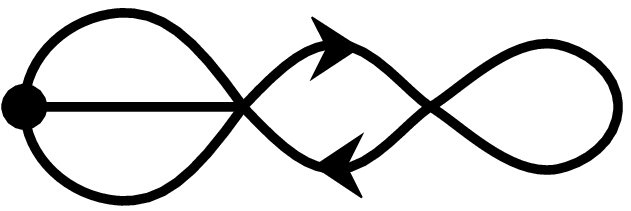}{70}{10} \notag\\
&\;+\;\frac{1}{8}\,\diagram{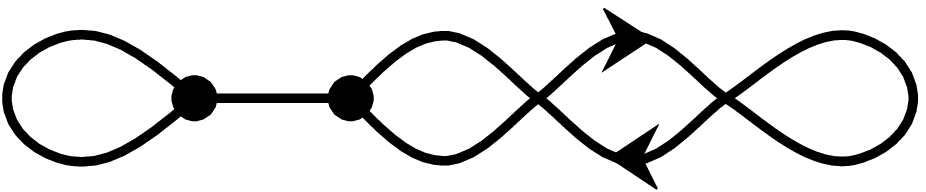}{95}{7}
\;+\;\frac{1}{16}\,\diagram{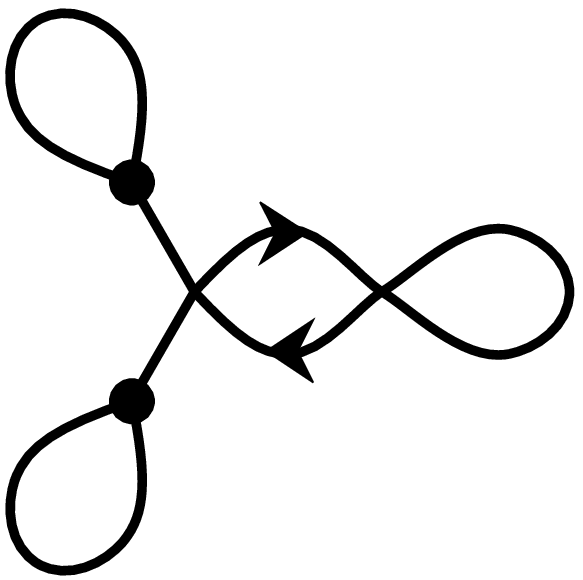}{60}{27}
\;+\;\frac{1}{8}\,\diagram{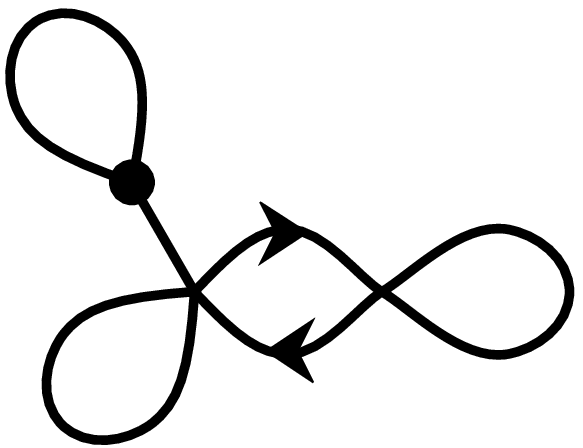}{60}{14}
\;+\;\frac{1}{4}\,\diagram{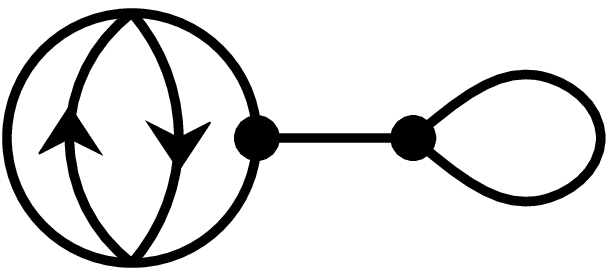}{65}{11}\notag\\
&\;+\;\frac{1}{4}\,\diagram{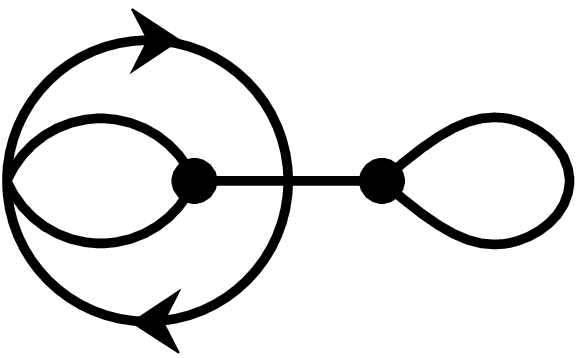}{65}{18} 
\;+\;\frac{1}{8}\,\diagram{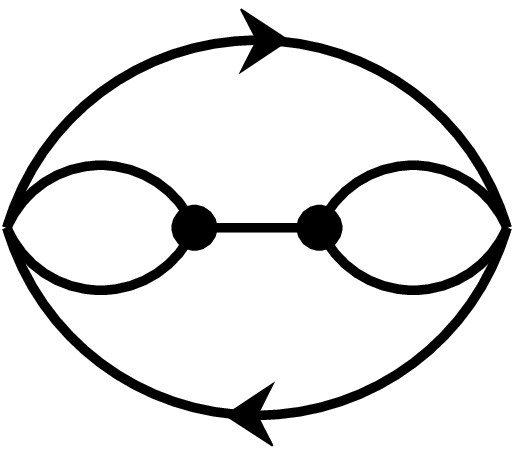}{50}{20}
\;+\;\frac{1}{8}\,\diagram{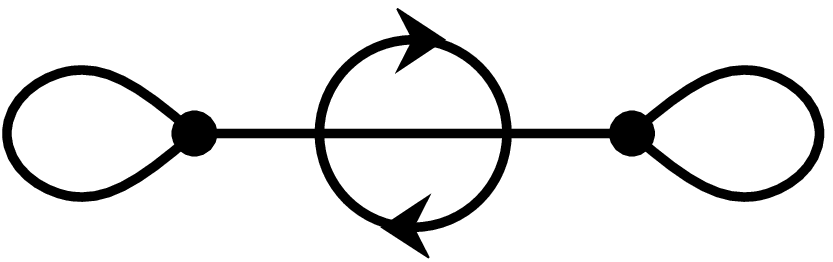}{85}{11}
\;+\;\frac{1}{4}\,\diagram{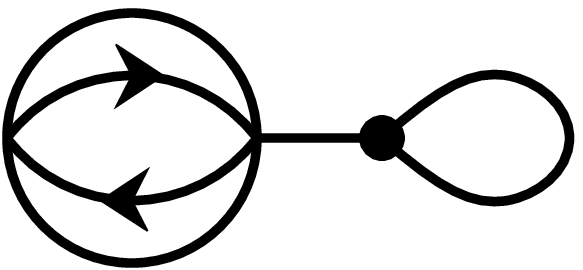}{65}{12}\notag \\
&\;+\;\frac{1}{4}\,\diagram{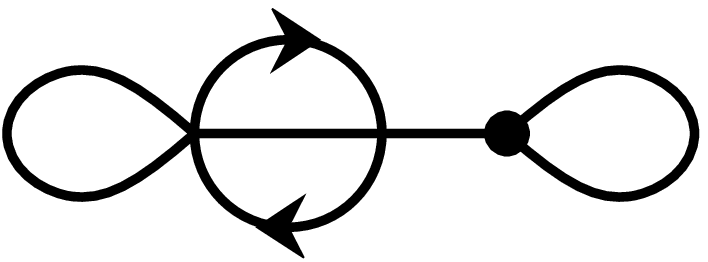}{75}{12} 
\;+\;\frac{1}{8}\,\diagram{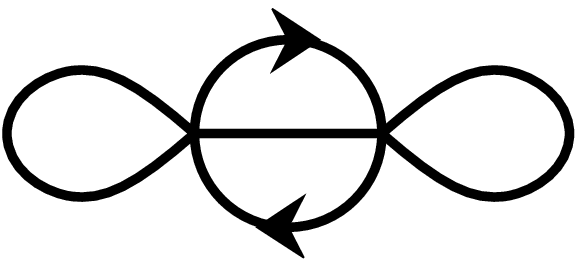}{60}{12}
\;+\;\frac{1}{4}\,\diagram{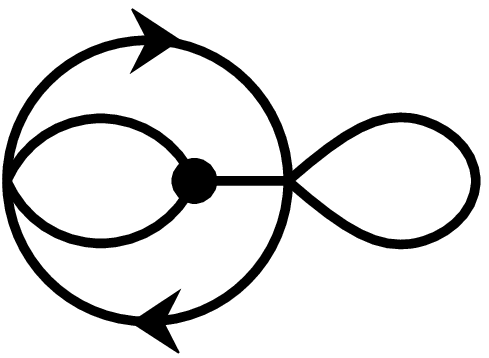}{55}{18}
\;+\;\frac{1}{3}\,\diagram{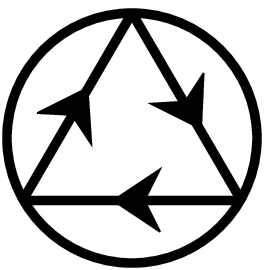}{30}{13}\notag \\
&\;+\;\frac{1}{24}\,\diagram{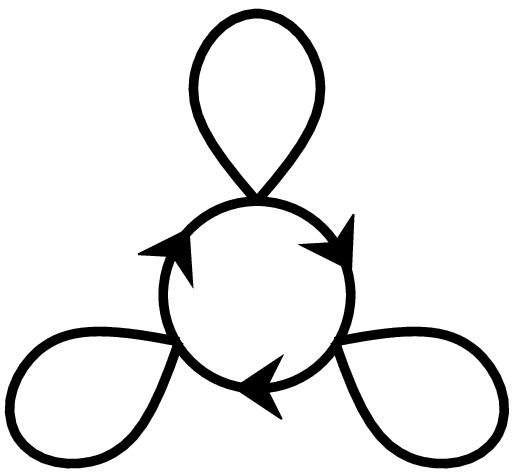}{54}{18}
\;+\;\frac{1}{4}\,\diagram{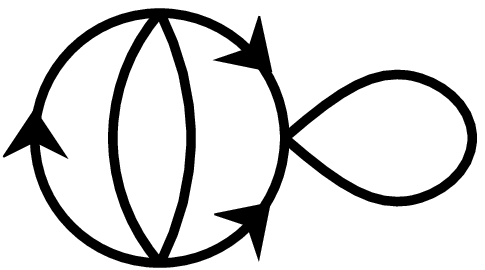}{50}{12}
\;+\;\frac{1}{16}\,\diagram{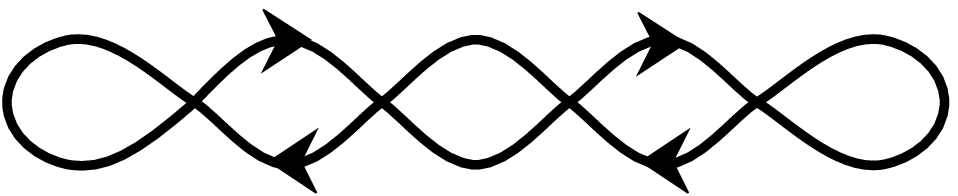}{90}{7}
\;+\;\frac{1}{4}\,\diagram{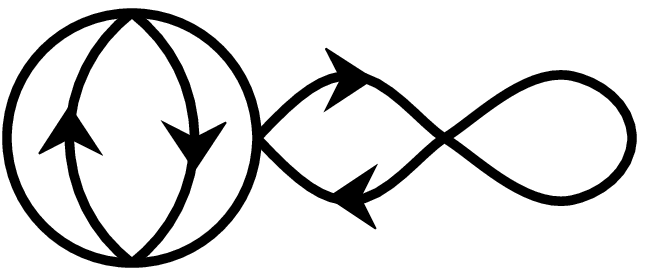}{75}{12}\notag
\end{align}

\subsection{One-vertex decomposability}
For some discrepancies, the contribution of a bosonic part of a diagram that
consists of two pieces connected by {\em only one} vertex, is equal to
the product of the contribution of those pieces. Such diagrams we call {\em
one-vertex reducible}, and discrepancies with this property we call {\em
one-vertex decomposable}. Examples of such discrepancies are those for which
$\twoB$ is translation invariant, i.e., $\twoB(x,y)=\twoB(x+a,y+a)$
$\forall\,x,y,a\in\Kube$, such as the Fourier diaphony. Also the Lego
discrepancy with equal bins is one-vertex decomposable. In contrast, the 
$L_2^*$-discrepancy is not one-vertex decomposable.

As a result of the one-vertex decomposability, many diagrams cancel or give 
zero. For example, the first and the second diagram in (\ref{CorEq009}) cancel, and the fourth gives zero, so that 
\begin{equation}
   \frac{1}{N}W_1(z) 
   \;=\;       \frac{1}{4}\,\diagram{dcA21}{30}{12}  
                \;+\; \frac{1}{12}\,\diagram{dcA12}{30}{11} \;\;.
\label{CorEq010}		
\end{equation}
To second order, only the following remains:
\begin{align}
\frac{1}{N^2}W_2(z) \;=\;
&    \frac{1}{48}\,\diagram{dcB12}{30}{11}
\;+\;\frac{1}{24}\,\diagram{dcB111}{30}{12}
\;+\;\frac{1}{8}\,\diagram{dcB18}{30}{12}
\;+\;\frac{1}{16}\,\diagram{dcB115}{40}{13} \notag\\
&\;+\;\frac{1}{8}\,\diagram{dcB24}{30}{16}  
\;+\;\frac{1}{4}\,\diagram{dcB28}{30}{16}  
\;+\;\frac{1}{4}\,\diagram{dcB29}{30}{12}
\;+\;\frac{1}{2}\,\diagram{dcB210}{30}{12}\notag\\
&\;+\;\frac{1}{16}\,\diagram{dcB36}{40}{15}
\;+\;\frac{1}{8}\,\diagram{dcB37}{35}{18}
\;+\;\frac{1}{4}\,\diagram{dcB38}{35}{15}
\;+\;\frac{1}{12}\,\diagram{dcB211}{30}{12} 
\;+\;\frac{1}{3}\,\diagram{dcB31}{30}{13} \;\;.
\label{CorEq015}
\end{align}

We now derive a general rule of diagram cancellation. First, we extend
the notion of one-vertex reducibility to complete diagrams, including the
fermionic part, with the rule that the two pieces both must contain a 
bosonic part. Consider the following diagram
\begin{equation}
   \diagram{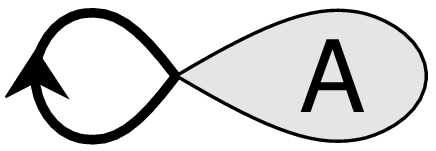}{65}{0}  \;\;.
\label{CorEq011}   
\end{equation}
The only restriction we put one the ``leave'' $A$ is that it must be one-vertex 
irreducible with respect to the vertex that connects it to the fermion loop.
For the rest, it may be anything. We define the contribution of the leave by
the contribution of the whole diagram divided by $-N$, and denote it with
$C(A)$. This contribution includes internal symmetry factors. Now consider a
diagram consisting of a fermion loop as in diagram (\ref{CorEq011}) with
attached to the one vertex $n_1$ leaves of type $A_1$, $n_2$ leaves of type
$A_2$, and so on, up to $n_p$ leaves of type $A_p$. The extra symmetry factor
of such a diagram is $(n_1!n_2!\cdots n_p!)^{-1}$, and, for one-vertex
decomposable discrepancies, the contribution is equal to the product of the
contributions of the leaves, so that the total contribution is given by
\begin{equation}
   -N\prod_{q=1}^p\frac{C(A_q)^{n_q}}{n_q!} \;\;.
\end{equation}
Now we sum the contribution of all possible diagrams of this kind that can
made with the $p$ leaves, and denote the result by
\begin{align}
   \diagram{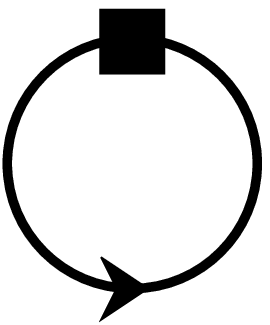}{30}{17} 
   \;=\; -N\sum_{n_1,n_2,\ldots\geq1}\prod_{q=1}^p
           \frac{C(A_q)^{n_q}}{n_q!} 
   \;=\; -N\left(e^{\sum_{q=1}^pC(A_q)} - 1\right) \;\;.
\label{CorEq012}   
\end{align}
Because the black square in l.h.s. of
\eqn{CorEq012} represents all possible ways to put the leaves together
onto one vertex, the sum of all possible ways to put the leaves onto one
fermion loop is given by
\begin{equation}
   \diagram{lcFl1}{30}{17} \;+\;
   \diagram{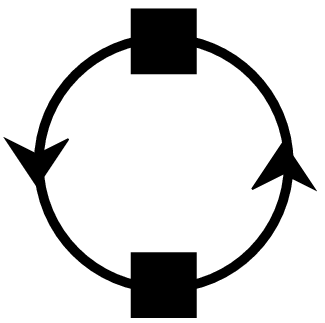}{36}{17} \;+\;
   \diagram{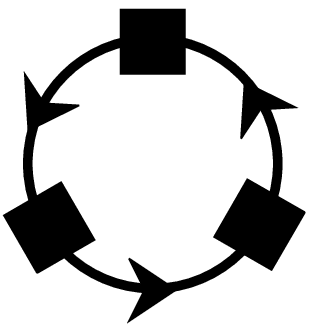}{35}{17} \;+\; \cdots 
   \;=\; -N\sum_{n=1}^{\infty}\frac{(-1)^{n-1}}{n}
         \left(e^{\sum_{q=1}^pC(A_q)} - 1\right)^n    \;\;.
\label{CorEq013}	 
\end{equation}
The $(-1)^{n-1}$ in the sum comes from the vertices and $1/n$ is the extra
symmetry factor of such diagram with $n$ vertices. The sum can be evaluated 
further and is equal to
\begin{equation}
    -N\log e^{\sum_{q=1}^pC(A_q)} \;=\; -N\sum_{q=1}^pC(A_q) \;\;,
\end{equation}
i.e., the sum of all possible ways to put $p$ different leaves onto one
fermion loop is equal to the sum of all leaves, each of them put onto its own 
fermion loop. This means that diagrams, consisting of two or more leaves put 
onto one fermion loop, cancel.

Now consider the following equation, which holds for 
every one-vertex decomposable discrepancy: 
\begin{equation}
   \diagram{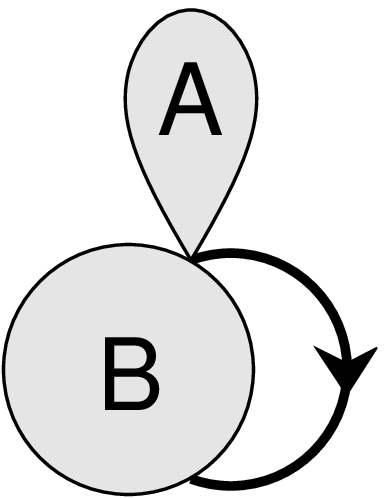}{50}{15} \;=\; -\; \diagram{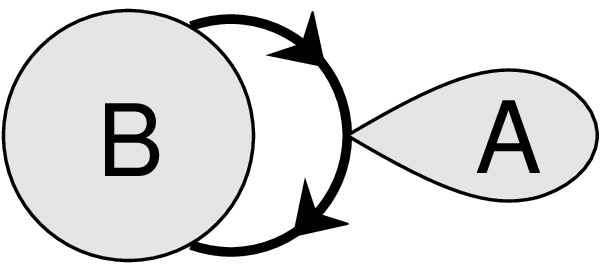}{80}{15} \;\;,
\label{CorEq014}   
\end{equation}
where we only assume that $B$ is not of the type on the l.h.s. of
\eqn{CorEq013}. The minus sign comes from the fact that the first diagram has
one vertex less. Because the number of fermion lines, a fermion loop consists
of is equal to the number of vertices it contains, we can always pair the
diagrams into one diagram of the l.h.s. type and one of the r.h.s. type so that
they cancel. We can summarize the result with the rule that {\em for one-vertex
decomposable discrepancies, only the one-vertex irreducible diagrams
contribute}.

\section{Applications}
We apply the general formulae given above to the Lego discrepancy, the 
$L_2^*$-discrepancy in one dimension and the Fourier diaphony in one dimension.

\subsection{The Lego discrepancy}
For the Lego problem class, for example defined in \cite{hak1}, the basis
consists of a set of characteristic functions $\vt_n$, $n=1,2,\ldots,M$ of $M$
disjunct subspaces of $\Kube$. We will denote the measure $\intk\vt_n(x)\,dx$
of subspace $n$ by $w_n$ and we have $\sum_{n=1}^Mw_n=1$. The strengths $\si_n$
are equal to $1/\sqrt{w_n}$\,, and the propagator is given by
\begin{equation}
   \twoB(x,y) \;=\; \sum_{n=1}^M\frac{\vt_n(x)\vt_n(y)}{w_n}-1 \;\;. 
\end{equation}
With this choice of $\si_n$, the discrepancy is just the $\chi^2$ statistic
that determines how well the points are distributed over the bins. It is easy
to see that $\twoB_p(x,y)=\twoB(x,y)$, $p=2,3,\ldots$, so that the
dressed propagator is given by 
\begin{equation}
   \prop_z(x,y)  \;=\; \frac{1}{1-2z}\,\twoB(x,y) \;\;.
\end{equation}
In \cite{hak1}, the propagator is given as an $M\times M$-matrix with matrix 
elements
\begin{equation}
   \prop^{(z)}_{n,m} \;=\; \frac{1}{1-2z}\left[\frac{\de_{n,m}}{w_n}-1\right]
\end{equation}
and acting in the $M$-dimensional function space, rather than as a two-point
function. This follows naturally from the path integral, which is an
$M$-dimensional integral. The obvious and correct relation between the two is
that 
\begin{equation}
   \prop_z(x,y) \;=\; \sum_{n,m=1}^M\vt_n(x)\prop^{(z)}_{n,m}\vt_m(y) \;\;.
\end{equation}
The zeroth order term can be found with the relation of \eqn{CorEq008}, which
results in the following expression
\begin{equation}
   W_0(z) \;=\; -\frac{1}{2}\log(1-2z)\intk\twoB(x,x)\,dx
               \;=\; -\frac{M-1}{2}\log(1-2z) \;\;,
\label{CorEq017}	       
\end{equation}
in agreement with Eq.~(44) in \cite{hak1}. To write down the first order term, 
we introduce 
\begin{equation}
   M_2 = \sum_{n=1}^M\frac{1}{w_n} \;\;,
   \quad\textrm{and}\quad \eta(z) = \frac{2z}{1-2z}  \;\;,
\end{equation}
so that 
\begin{equation}
   W_1(z)
   \;=\; \frac{1}{8}\left(M_2-M^2-2M+2\right)\eta(z)^2
         +\frac{1}{24}\left(5M_2-3M^2-6M+4\right)\eta(z)^3  \;\;.
\label{CorEq018}	 
\end{equation}
If the bins are equal, so that $w_n=1/M$ $n=1,2,\ldots,M$, then only the 
contribution of the diagrams of \eqn{CorEq011} remains, and the result is 
\begin{equation}
   W_1(z)
   \;=\; -\frac{1}{4}\,E\eta(z)^2 + \frac{1}{12}(E^2-E)\eta(z)^3 \;\;,
\label{CorEq019}   
\end{equation}
where we denote
\begin{equation}
   E = M-1 \;\;.
\end{equation}
To second order in $1/N$, the contribution comes from the diagrams in \eqn{CorEq015}, and is given by 
\begin{align}
   W_2(z)
   \;=\;       &(5E^3-12E^2+7E)\frac{\eta(z)^6}{48} 
         \;+\;  (E^3-6E^2+5E)\frac{\eta(z)^5}{8}    \notag\\
	 \;&+\; (E^3-28E^2+43)\frac{\eta(z)^4}{48}
	 \;+\;  (-E^2-5E)\frac{\eta(z)^3}{12}       \;\;.
\label{CorEq020}   
\end{align}
In Appendix A, we present the expansion of $G(z)$ in the case of equal bins, up
to and including the $1/N^4$ term. It is calculated using the path integral
expression (\ref{CorEq002}) of $G(z)$ and computer algebra. The reader may
check that this expression for $G(z)$ and the above terms of $W(z)$ satisfy
$G(z)=e^{W(z)}$ up to the order of $1/N^2$.

\subsection{The $\boldsymbol{L}_{\boldsymbol{2}}^{\boldsymbol{*}}$-discrepancy}
For the $L_2^*$-discrepancy in one dimension, for example defined in
\cite{hak1}, the gauge freedom is a freedom in the boundary conditions which the
members of the problem class have to satisfy. $\vLa$ acts on the members as
\begin{equation}
   (\vLa\phi)(x) \;=\; -\frac{d^2\phi}{dx^2}(x) \;\;, 
\label{CorEq111}   
\end{equation}
and in the Landau gauge, the boundary conditions are given by
\begin{equation}
   \intk\phi(x)\,dx = 0 \;\;,\quad 
   \frac{d\phi}{dx}(0)=\frac{d\phi}{dx}(1)=0\;\;.
\end{equation}
The basis in the Landau gauge is given by the set of eigenfunctions of $\vLa$
with the boundary conditions above, which is 
$\{\sqrt{2}\cos(n\pi x),\; n=1,2,\ldots\}$,  
so that the propagator is given by 
\begin{align}
   \twoB(x,y) 
   \;=\; \sum_{n=1}^\infty\frac{2\cos(n\pi x)\cos(n\pi y)}{n^2\pi^2} 
   \;=\; \min(x,y) - x + \half x^2 - y + \half y^2 + \sfrac{1}{3}\;\;.
\end{align}
The dressed propagator is given by 
\begin{align}
   \prop_z(x,y) 
   \;&=\; \sum_{n=1}^\infty\frac{2\cos(n\pi x)\cos(n\pi y)}{n^2\pi^2-2z} \\
   \;&=\; \frac{1}{u^2} - \frac{1}{2u\sin u}
                          \{\cos[u(1-|x+y|)] \;+\; \cos[u(1-|x-y|)]\} \;\;,
\end{align}
with 
\begin{equation}
   u=\sqrt{2z}  \;\;.
\end{equation}
The zeroth order term can be obtained using \eqn{CorEq008}: 
\begin{equation}
   W_0(z) \;=\; -\frac{1}{2}\log\left(\frac{\sin u}{u}\right) \;\;,
\label{CorEq021}   
\end{equation}
which is the well-known result. After some algebra, also the first order term
follows:
\begin{equation}
   W_1(z) 
   \;=\; \frac{1}{288}\left(24-8\frac{u}{\sin u}-7\frac{u^2}{\sin^2u}
                          -7\frac{u}{\tan u}-2\frac{u^2}{\tan^2u}\right) \;\;.
\label{CorEq022}   
\end{equation}

\subsection{The Fourier diaphony}
Usually, the Fourier diaphony is defined in terms of a basis that is in the 
Landau gauge already. It is, for example, given in \cite{hkh1}, and in one 
dimension, the basis is given by the functions 
\begin{equation}
   u_{2n-1}(x)=\sqrt{2}\sin(2\pi nx) \;\;,\quad
   u_{2n}(x)  =\sqrt{2}\cos(2\pi nx) \;\;,\quad n=1,2,\ldots \;\;,
\end{equation}   
and for the strengths we take
\begin{equation}
   \si_{2n-1} = \si_{2n} = \frac{1}{n} \;\;,\quad n=1,2,\ldots \;\;. 
\end{equation}
Notice that the basis functions satisfy
$-\frac{d^2}{dx^2}u_n(x)=\frac{4\pi^2}{\si_n^2}u_n(x)$, so that, from this
point of view, the only relevant difference between the $L_2^*$-discrepancy and
the Fourier diaphony are the boundary conditions on the members of the problem
class. 

The propagator is given by 
\begin{align}
   \twoB(x,y) 
   \;=\; \sum_{n=1}^\infty\frac{2\cos(2n\pi\{x-y\})}{n^2} 
   \;=\; \frac{\pi^2}{3}\left[1-6\{x-y\}(1-\{x-y\})\right]  \;\;,
\end{align} 
where we use the notation $\{x\}=x\mod1$. The dressed propagator is given by 
\begin{equation}
   \prop_z(x,y)
   \;=\; \sum_{n=1}^\infty\frac{2\cos(2n\pi\{x-y\})}{n^2-2z}
   \;=\; \frac{\pi^2}{v^2}\left(1-\frac{v\cos[v(2\{x-y\}-1)]}{2\sin v}\right)
   \;\;,
\end{equation}
where
\begin{equation}
   v=\sqrt{2\pi^2 z} \;\;.
\end{equation}
This two-point function is, apart of a factor $\pi^2/v^2$, the same as the one 
in Eq.~(26) in \cite{hk2}.
The zeroth order term can easily be obtained from the dressed propagator and 
is given by
\begin{equation}
   W_0(z) \;=\; -\log\left(\frac{\sin v}{v}\right) \;\;,
\label{CorEq023}   
\end{equation}
which is in correspondence with Eq.~(21) in \cite{hk2}.
Because the propagator is translation invariant, i.e.,
$\twoB(x+a,y+a)=\twoB(x,y)$ $\forall\,x,y,a\in\Kube$, the contributions
of the first two diagrams in \eqn{CorEq009} cancel, and the contribution of the
fourth diagram is zero. The contribution of the remaining diagrams gives
\begin{equation}
   W_1(z) 
   \;=\; \frac{1}{36}\left(3 + v^2 - 3\frac{v^2}{\sin^2v}\right) \;\;.
\label{CorEq024}   
\end{equation}

\section{Conclusions}
In addition to the machinery of QFT introduced in \cite{hak1}, we introduced
fermions to calculate the moment generating function $G$ of the probability
distribution under sets of random points of a quadratic discrepancy $D_N$.
They allow for an expansion in the inverse of the number of points $N$ of the
logarithm $W$ of $G$, where the contribution to each term in the expansion can
be represented by a finite number of connected Feynman diagrams. We have
presented the diagrams up to the order of $1/N^2$ for the general case, and
derived a rule of diagram cancellation in the case of special discrepancies,
which we call one-vertex decomposable.

We have applied the formalism to the Lego discrepancy, the $L_2^*$-discrepancy
in one dimension and the Fourier diaphony in one dimension, and calculated the
first two terms $W_0$ and $W_1/N$ in the expansion. For the Lego discrepancy,
this resulted in \eqn{CorEq017} and \eqn{CorEq018}, for the $L_2^*$-discrepancy
in \eqn{CorEq021} and \eqn{CorEq022}, and for the Fourier diaphony in
\eqn{CorEq023} and \eqn{CorEq024}. The Fourier diaphony and the Lego
discrepancy with equal binning are one-vertex decomposable. For the latter, we
also calculated the term $W_2/N^2$, which is in correspondence with the result
of an alternative calculation up to the order of $1/N^4$, given in Appendix A.

Calculations become very cumbersome for high orders because of 
the number of diagrams involved. A situation in which the formalism can still 
be powerful is when another parameter in the definition of the discrepancy, 
such as the dimension of the integration region or the number of bins 
in case of the Lego discrepancy, becomes large. This parameter can then serve 
as an extra order parameter in the determination of the importance of the 
contribution of the diagrams, and can lead to a substantial reduction in the 
number of relevant diagrams. In \cite{hak4}, we will present our results 
with respect to this for the Lego discrepancy.

\section*{Appendix A}
If we define, for the Lego-discrepancy with equal bins, $E=M-1$,
$\eta(z)=2z/(1-2z)$ and 
\begin{equation}
   (1-2z)^{E/2}G(z) 
   \;=\; \sum_{n,p\geq0}\frac{\eta(z)^p}{N^n}\,C^{(p)}_n(E) \;\;,
\end{equation}
then the only non-zero $C^{(p)}_n(E)$ up to $n=4$ are given by 
\begin{align}
   C_1^{(2)}(E) &= -\frac{1}{4}E\notag\\
   C_1^{(3)}(E) &= E\bigg(\frac{1}{12}E-\frac{1}{12}\bigg)\notag\\
   C_2^{(3)}(E) &= E\bigg(-\frac{1}{12}E+\frac{5}{12}\bigg)\notag\\
   C_2^{(4)}(E) &= E\bigg(\frac{1}{48}E^2-\frac{53}{96}E
                   +\frac{43}{48}\bigg)\notag\\
   C_2^{(5)}(E) &= E\bigg(\frac{5}{48}E^2-\frac{35}{48}E
                   +\frac{5}{8}\bigg)\notag\\
   C_2^{(6)}(E) &= E\bigg(\frac{1}{288}E^3+\frac{7}{72}E^2
                   -\frac{71}{288}E+\frac{7}{48}\bigg)\notag\\
   C_3^{(4)}(E) &= E\bigg(-\frac{1}{48}E^2+\frac{7}{12}E
                   -\frac{61}{48}\bigg)\notag
\end{align}
\begin{align}
   C_3^{(5)}(E) &= E\bigg(\frac{1}{240}E^3-\frac{17}{30}E^2
                   +\frac{583}{120}E-\frac{1451}{240}\bigg)\notag\\
   C_3^{(6)}(E) &= E\bigg(\frac{53}{576}E^3-\frac{1153}{384}E^2
                   +\frac{7423}{576}E-\frac{527}{48}\bigg)\notag\\
   C_3^{(7)}(E) &= E\bigg(\frac{1}{576}E^4+\frac{461}{1152}E^3
                   -\frac{6581}{1152}E^2+\frac{8663}{576}E
                   -\frac{467}{48}\bigg)\notag\\
   C_3^{(8)}(E) &= E\bigg(\frac{11}{1152}E^4+\frac{85}{144}E^3
                   -\frac{5125}{1152}E^2+\frac{1555}{192}E
                   -\frac{17}{4}\bigg)\notag\\
   C_3^{(9)}(E) &= E\bigg(\frac{1}{10368}E^5+\frac{29}{3456}E^4
                   +\frac{955}{3456}E^3-\frac{12475}{10368}E^2
                   +\frac{953}{576}E-\frac{53}{72}\bigg)\notag\\
   C_4^{(5)}(E) &= E\bigg(-\frac{1}{240}E^3+\frac{37}{80}E^2
                   -\frac{337}{80}E+\frac{1397}{240}\bigg)\notag\\
   C_4^{(6)}(E) &= E\bigg(\frac{1}{1440}E^4-\frac{349}{960}E^3
                   +\frac{7193}{720}E^2-\frac{15283}{320}E
                   +\frac{67021}{1440}\bigg)\notag\\
   C_4^{(7)}(E) &= E\bigg(\frac{49}{960}E^4-\frac{29069}{5760}E^3
                   +\frac{372169}{5760}E^2-\frac{571727}{2880}E
                   +\frac{21503}{144}\bigg)\notag\\
   C_4^{(8)}(E) &= E\bigg(\frac{13}{23040}E^5+\frac{13979}{23040}E^4
                   -\frac{2290601}{92160}E^3+\frac{1446743}{7680}E^2\notag\\
                &\phantom{= E\bigg(\frac{13}{23040}E^5}
		   -\frac{9583187}{23040}E+\frac{294773}{1152}\bigg)\notag\\
   C_4^{(9)}(E) &= E\bigg(\frac{73}{6912}E^5+\frac{35077}{13824}E^4
                   -\frac{781079}{13824}E^3+\frac{993515}{3456}E^2
                   -\frac{564301}{1152}E+\frac{24607}{96}\bigg)\notag\\
   C_4^{(10)}(E) &= E\bigg(\frac{1}{13824}E^6+\frac{139}{3072}E^5
                    +\frac{162721}{34560}E^4-\frac{596467}{9216}E^3\notag\\
		 &\phantom{= E\bigg(\frac{1}{13824}E^6}   
                    +\frac{1653251}{6912}E^2-\frac{253799}{768}E
                    +\frac{145199}{960}\bigg)\notag\\
   C_4^{(11)}(E) &= E\bigg(\frac{17}{41472}E^6+\frac{895}{13824}E^5
                    +\frac{55025}{13824}E^4-\frac{1505645}{41472}E^3\notag\\
		 &\phantom{= E\bigg(\frac{17}{41472}E^6}   
                    +\frac{19783}{192}E^2-\frac{137875}{1152}E
                    +\frac{1565}{32}\bigg)\notag\\
   C_4^{(12)}(E) &= E\bigg(\frac{1}{497664}E^7+\frac{11}{31104}E^6
                    +\frac{2431}{82944}E^5+\frac{155735}{124416}E^4\notag\\
		 &\phantom{= E\bigg(\frac{1}{497664}E^7}   
                    -\frac{3942431}{497664}E^3+\frac{249239}{13824}E^2
                    -\frac{250141}{13824}E+\frac{2575}{384}\bigg)\notag
\end{align}

\end{document}